\documentclass[aps,prb,twocolumn,superscriptaddress,showpacs,floatfix,longbibliography]{revtex4-1}
\usepackage{amsmath,amssymb}

\usepackage[utf8]{inputenc}
\usepackage[T1]{fontenc}
\usepackage{xcolor}
\usepackage{booktabs}
\usepackage{array}
\IfFileExists{newtxtext.sty}
   {\usepackage{newtxtext,newtxmath}}
   {\IfFileExists{stix.sty}
      {\usepackage{stix}}
      {\IfFileExists{mathptmx.sty}
      {\usepackage{mathptmx}}{} } }

\usepackage{textcomp}

\usepackage{bm}

\IfFileExists{siunitx.sty}{\usepackage{booktabs,siunitx}}{}

\pdfoutput=1
\usepackage{color}
\definecolor{LinkColor}{rgb}{0.256,0.439,0.588}
\usepackage{hyperref}
\hypersetup{
   pdfauthor={abc},
   pdftitle={paper},
   colorlinks=true,
   citecolor=blue,
   linkcolor=blue,
   urlcolor=blue
}

\usepackage{pifont}

\usepackage[pdftex]{graphicx}
\DeclareGraphicsExtensions{.pdf,.jpeg,.png,.eps}
\usepackage{epstopdf}
\usepackage{soul}

\begin{document}
	\bibliographystyle{apsrev4-1}
	\title{Violation of Luttinger's theorem in the simplest doped Mott insulator: Falicov-Kimball model in strong correlation limit}
	
	\author{Wei-Wei Yang}
	\affiliation{School of Physical Science and Technology $\&$ Key Laboratory for Magnetism and Magnetic Materials of the MoE, Lanzhou University, Lanzhou 730000, People Republic of China} %
	\author{Qiaoni Chen}
	\affiliation{Department of Physics, Beijing Normal University, Beijing 100875, People Republic of China}
	\author{Hong-Gang Luo}
	\affiliation{School of Physical Science and Technology $\&$ Key Laboratory for Magnetism and Magnetic Materials of the MoE, Lanzhou University, Lanzhou 730000, People Republic of China} %
	\affiliation{Beijing Computational Science Research Center, Beijing 100084, China}%
    \affiliation{Lanzhou Center for Theoretical Physics, Key Laboratory of Theoretical Physics of Gansu Province}

	\author{Yin Zhong}
    \email{zhongy@lzu.edu.cn}
    \affiliation{School of Physical Science and Technology $\&$ Key Laboratory for Magnetism and Magnetic Materials of the MoE, Lanzhou University, Lanzhou 730000, People Republic of China} %
    \affiliation{Lanzhou Center for Theoretical Physics, Key Laboratory of Theoretical Physics of Gansu Province}
	
\begin{abstract}	
The Luttinger's theorem has long been taken as the key feature of Landau's Fermi liquid, which signals the presence of quasiparticles. Here, by the unbiased Monte Carlo method, violation of Luttinger's theorem is clearly revealed in the Falicov-Kimball (FK) model, indicating the robust correlation-driven non-Fermi liquid characteristic under any electron density. Introducing hole carriers to the half-filled FK leads to Mott insulator-metal transition, where the Mott quantum criticality manifests unconventional scaling behavior in transport properties.
Further insight on the violation of the Luttinger's theorem is examined by combining Hubbard-I approximation with a composite fermion picture, which emphasizes the importance of a mixed excitation of the itinerant electron and the composite fermion. Interestingly, when compared FK model with a binary disorder system, it suggests that the two-peak band structure discovered by Monte Carlo and Hubbard-I approaches is underlying the violation of Luttinger's theorem.
\end{abstract}
	
	\date{\today}
	
	\maketitle
	
	
\section{\label{sec1:level1}Introduction}	
Electron correlation has long been the key ingredient of modern condensed matter physics. In strongly correlated systems, such as the heavy fermion compound, cuprates and the iron-based superconductors, correlation introduces various unconventional metallic states \cite{RevModPhys.78.17,RevModPhys.83.1589,Powell_2011}. The metallic state with linear-$T$ resistivity is often called strange metal after its discovery in the normal state of high-$T_{c}$ cuprates superconductors. Similar phenomena has also been found in heavy fermion compounds where magnetic quantum critical point is approaching. These unconventional metallic states cannot be understood in the framework of Landau's Fermi liquid (FL) theory and has been empirically classified as non-Fermi liquids (NFLs) due to anomalous thermodynamic and transport properties\cite{PhysRevB.104.235138}. Unfortunately, despite intensive decades of studies, our knowledge on generic NFLs (beyond artificial large-$N$ limits or solvable models) is still very limited because the intrinsic strong correlation effect is beyond the scope of Hartree-Fock mean-field and perturbation theory framework, thus intuitive understanding of these NFLs is largely unknown.

Fortunately, the Luttinger's theorem provides a heuristic way to understand the essence of the NFLs. The Luttinger's theorem suggests the existence of quasiparticles, and has long been taken as a key feature of FL. Recently, several numerical researches have confirmed that the Luttinger's theorem is violated in the doped correlation-driven Mott insulators, e.g., in Fermi-Hubbard model, $t-V$ model and $t-J$ model \cite{landau1957theory,PhysRevB.104.235122,huang2019strange,PhysRevB.54.4336,PhysRevB.78.153103,PhysRevLett.81.2966,PhysRevB.75.045111,PhysRevB.62.4336,Kokalj63,Ortloff58}. However, the accuracy of these studies is affected by the dimension, size of the system, limited temperature region, and crude approximation. Especially for a doped Mott insulator lacking of particle-hole symmetry, due to the notorious minus-sign problem, the most trustworthy determinant quantum Monte Carlo simulation of the Fermi-Hubbard model is limited to a small size and high temperature, whereas the analytical study can rarely predict the nature of a specific system. In our previous work, we have studied the violation of the Luttinger's theorem in the anisotropic limit of Kondo lattice, i.e., the Ising-Kondo lattice model \cite{PhysRevB.104.165146}. Considering the specific feature of Ising-Kondo lattice, the unbiased study of the violation of the Luttinger's theorem is still insufficient and a more generic and transparent system which permits unbiased Monte Carlo simulation on a large size close to the thermodynamic limit is heavily desired..

We note that the Falicov-Kimball (FK) model is such an ideal platform to study the Luttinger's violation of the doped Mott insulator. The FK model is an alternative Hubbard model, which can be exactly solved within the framework of dynamical mean-field theory (DMFT) in infinite dimension \cite{RevModPhys.75.1333}. Hitherto, most research about FKM are restricted to the weak coupling regime, where the long-range correlation and the natural tendency to phase separation conspire to rich patterns in the zero-temperature phase diagram. \cite{PhysRevB.74.035109,RevModPhys.75.1333} These stripelike or inhomogeneous charge orderings can correspond to kinds of strongly correlated electron systems \cite{DAGOTTO20011,Tokura_2006,vojta2009lattice}, such as the doped cuprates \cite{tranquada1995evidence,Berg_2009,parker2010fluctuating,wu2011magnetic,abbamonte2005spatially,PhysRevB.78.174529,Fausti189,Ghiringhelli821,Comin1335,PhysRevLett.120.037003,zhao2019charge}, layered cobalt oxides \cite{boothroyd2011hour,babkevich2016direct} and nickelates \cite{PhysRevLett.104.206403,PhysRevLett.122.247201,Norman_2016,zhang2017large,Cosloviche1600735}. It is used to be thought that the FK model in small interaction (use $U$ as its strength) is more unconventional, while the large $U$ situation can be simply mapped onto effective Ising model and thus its underlying physics has been rarely reported in literature\cite{https://doi.org/10.1002/pssb.200460067}. In this paper we instead focus on the strong coupling regime and reveal that the strong correlation leads to non-trivial quantum states beyond Landau's FL paradigm. At high temperature, the FK system has no long-ranged charge order and thus could be clearly studied in terms of the violation of the Luttinger's theorem. It was reported that the Anderson insulator can be induced by any finite correlation at the thermodynamic limit \cite{PhysRevLett.117.146601}. Here, we find that in the doped Anderson insulator with sufficient correlation ($U>U_c$), single band evolves into a robust two-band structure, with which the Luttinger's theorem is violated. The disappearance of quasiparticle suggests the crossover from FL to NFL occurs in the Anderson insulator with increasing interaction. The evolution of band structure and the Luttinger integral can be exactly covered by a binary disordered system, with which we confirm the two-band structure is underlying the violation of Luttinger's theorem.
Actually, not all the NFLs can violate the Luttinger's theorem.
When Mott gap is absence, where the single-band structure in the NFL sustains under interaction, such as the Sachdev-Ye-Kitaev model, the Luttinger's theorem is satisfied \cite{SYK_Luttinger}.
Combined with previous studies, for the specific NFLs arising from the doped Mott insulator, the NFL nature is attributed to the change of band structure, and this kind of NFLs can be confirmed by the violation of the Luttinger's theorem.


The remainder of this paper is organized as follows: In Sec.~\ref{sec2}, the FK model is introduced and its phase diagram is briefly discussed. Strange metal state is demonstrated by the calculation of resistivity and specific heat. In Sec.~\ref{sec3}, we show that the Luttinger's theorem is violated in the FK model in terms of direct numerical calculation. In Sec.~\ref{sec4}, we provide some analytical result with the Hubbard-I approach and analyses it by the composite fermion. In Sec.~\ref{sec5}, with the assistance of the Luttinger's theorem, the phase diagram in the $U-T$ plane at half filling is elaborated. A binary disordered system is discussed. We compare the violation of the Luttinger's theorem and the spectrum function between the disordered system and the FK model, and finally end the article with a summary.


\section{Model and Method}\label{sec2}
We consider the FK model \cite{RevModPhys.75.1333} on a square lattice, the Hamiltonian is defined as
\begin{equation}
	\hat{H}=-t\sum_{i,j}\hat{c}_{i}^{\dag}\hat{c}_{j}+
	U\sum_{i}\hat{n}_{i}\hat{w}_{i}-\mu\sum_{i}\hat{n}_{i},
	\label{eq:model1}
\end{equation}
where $\hat{c}_{j}^{\dag}(\hat{c}_{j})$ is the itinerant electron's creation (annihilation) operator at site $j$. $\hat{n}_{j}=\hat{c}_{j}^{\dag}\hat{c}_{j}$ denotes the particle number of the itinerant electron, while $\hat{w}_j$ denotes the particle number operator for the electron in the localized state. $t$-term denotes the hopping integral and only nearest-neighbor-hoping is involved. $U$ is the onsite Coulomb interaction between the itinerant electron and the localized electron. The hole doping into the half-filled system ($\mu=\frac{U}{2}$) is realized by tuning the chemical potential $\mu$.

In Eq.~\ref{eq:model1} the number of localized electron at each site is conservative since $[\hat{w}_{j},\hat{H}]=0$.
Therefore, taking the eigenstates of the number operator of localized electron $\hat{w}_{j}$ as bases, the Hamiltonian is automatically reduced to an effective free fermion model
\begin{equation}
	\hat{H}(w)=-t\sum_{i,j}\hat{c}_{i}^{\dag}\hat{c}_{j}+
	U\sum_{i}{n}_{i}\hat{w}_{i}-\mu\sum_{i} {n}_{i}.
	\label{eq:eff_model1}
\end{equation}
Here, $w$ emphasizes its $w$ dependence and $\hat{w}_{i}|w_i>=w_i|w_i>$, $w_i=0,1$.
Now the many-body eigenstate of the original model can be constructed by the single-particle state of the effective Hamiltonian under a given configuration $\{w_i\}$,
and thus the Monte Carlo simulation can be simply carried out.
In the context we consider the square lattice with periodic boundary conditions. The sampling using the truncation algorithm is used, with which we are able to carry on a simulation of the system as large as $N_s=3600$ \cite{Kumar2006}. Since the thermodynamic and transport properties of FK model has been studied in literature, here we focus on the nature of Fermi surface. Therefore, a $60 \times 60$ square lattice is mainly used in the calculation of Luttinger integral, while the others are calculated with a $20 \times 20$ square lattice. Accordingly, the two-dimensional Brillouin zone are sampled by a $60\times 60$ $k$-point grid. Nearest neighbor hopping integral is used as the unit ($t=1$) to measure all energy scales. To attack the NFLs in the doped Mott insulator, we focus on the strong coupling regime ($U=10$).
In this work, we dope the half-filled FK model and consider the hole-doping case by tuning the chemical potential $\mu$. The translation from hole doping to electron doping is straightforward due to the particle-hole symmetry. Compared with Hubbard model, due to the lack of spin degree of freedom, in FK model the itinerant electron on each site contains only one possible state. The particle-hole symmetry also dictates that the particle density is $n_c=0.5$ when $\mu=\frac{U}{2}$.

\begin{figure}
	\centering
	\includegraphics[width=0.95\columnwidth]{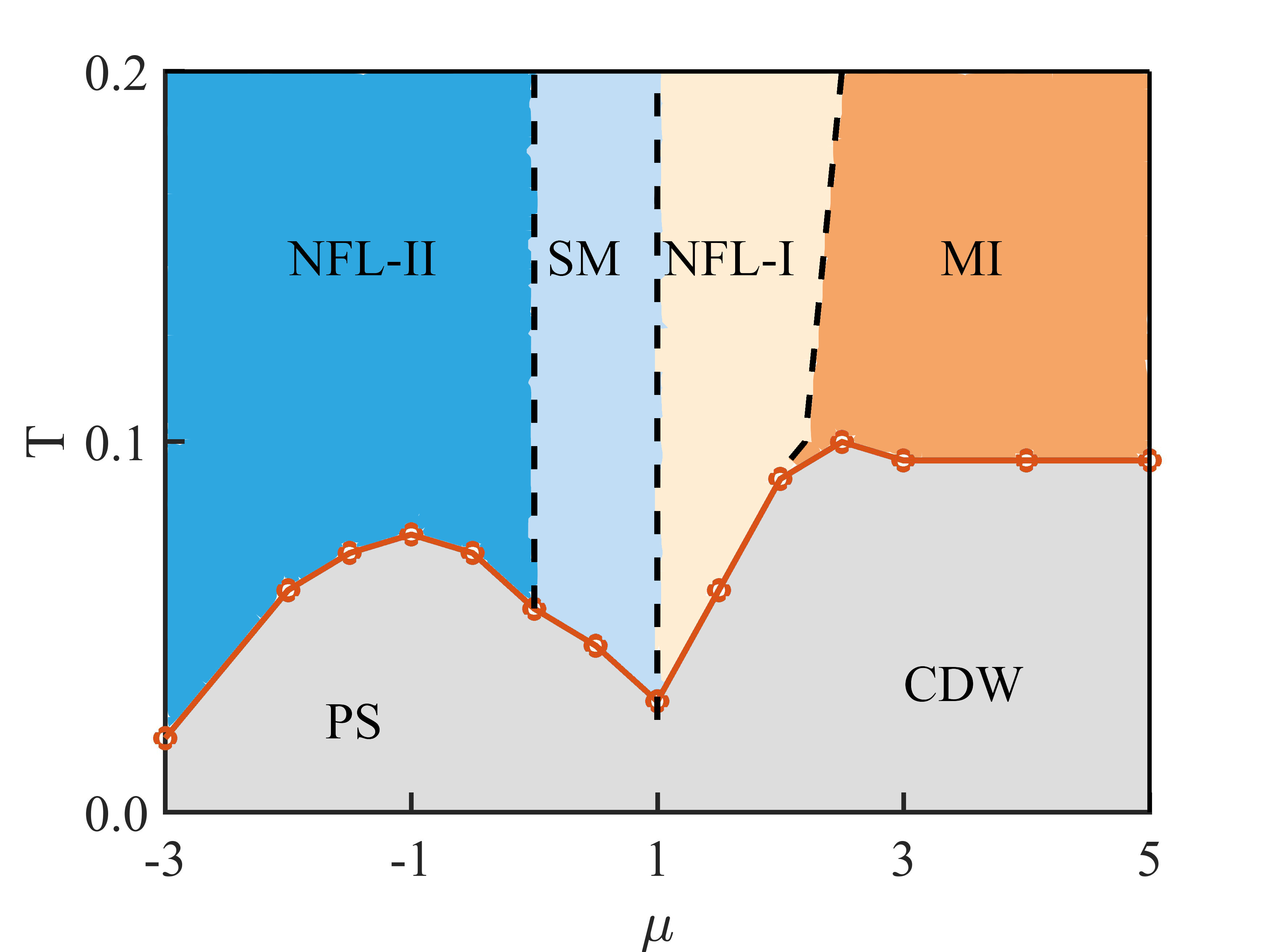}
	\caption{The phase diagram of the Falicov-Kimball (FK) model in the $\mu-T$ plane at $U=10$. The low temperature regime is divided into a charge density wave (CDW) state and a phase separation (PS) state by a first-order transition. At high temperature, there exists a Mott insulator state around half-filling situation and three distinguishable non-Fermi liquid (NFL) states. A quantum critical region is induced by doping-driven Mott insulator-metal transition. With increasing doped holes, the Mott insulator crossover successively to the first non-Fermi liquid (NFL-I) state, the strange metal (SM) state, and finally to the second non-Fermi liquid (NFL-II) state.}
	\label{fig:phase diagram}
\end{figure}

\begin{figure}
		\centering
		\includegraphics[width=1\columnwidth]{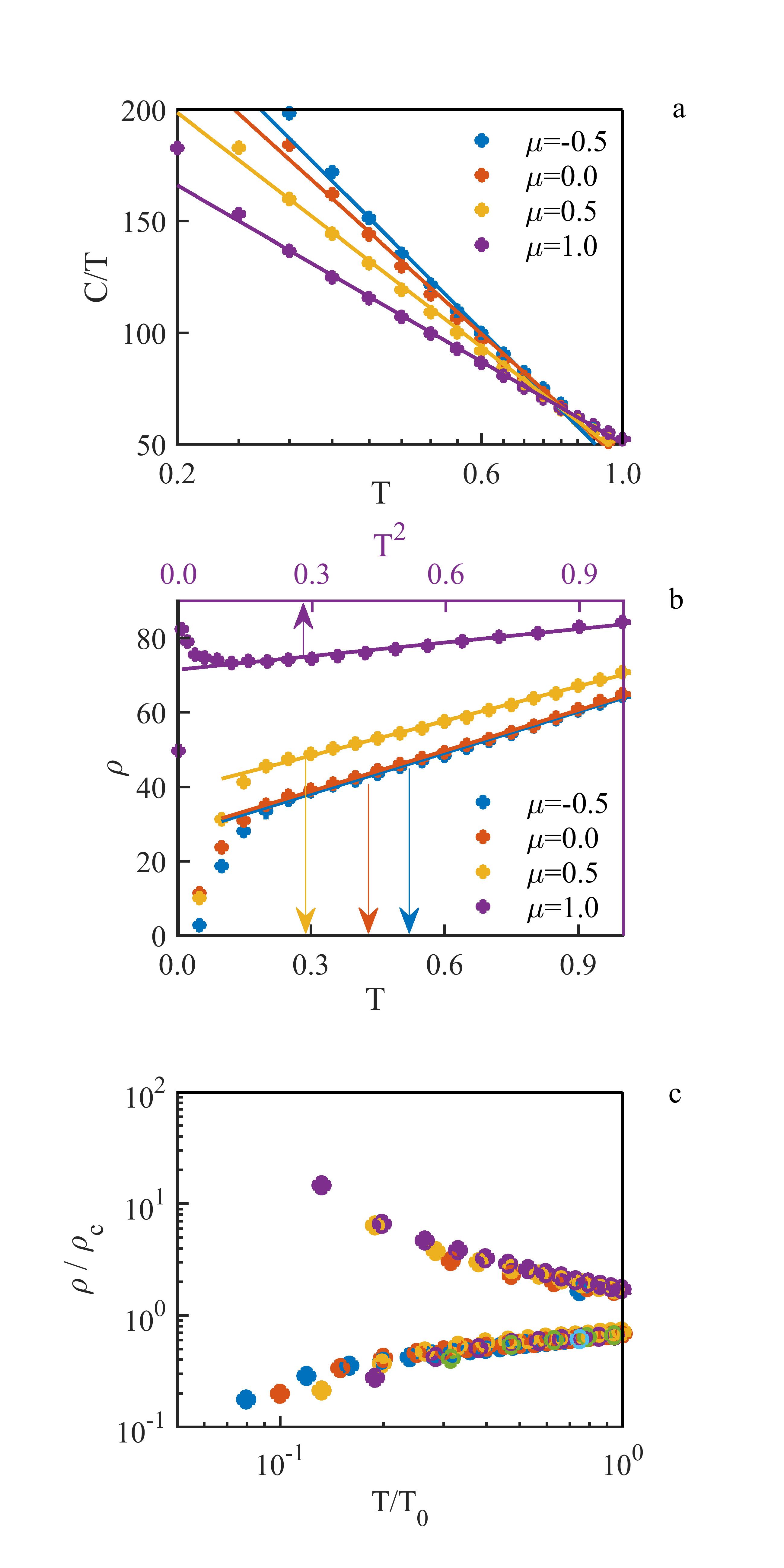}
		\caption{The SM behavior in the FK model. (a) Logarithm temperature dependence of the capacity heat coefficient $C(T)/T$. (b) The linear-$T$ resistivity. (c) The quantum critical scaling behavior. In the quantum critical region the resistivity satisfies the quantum critical scaling. The SM is the metal-like part and the NFL-I is referred to as the insulatorlike part.}
		\label{fig:Cv}
	\end{figure}

As indicated in the previous studies in Refs.\cite{PhysRevB.74.035109,doi:10.1080/01411594.2011.604509,https://doi.org/10.1002/pssb.200460067}, the low-temperature ordered states could be identified with structure factor $S_q^{\omega}(Q)$, correlation function $g_n$ and the renormalized correlation function $G_n$, which are respectively defined as :
\begin{equation}
	\begin{aligned}
	&g_n=\frac{1}{4N}\sum_{i=1}^{N}\sum_{\tau_1,\tau_2=\pm1}w(\textit{\textbf{r}}_i)w(\textit{\textbf{r}}_i+\tau_1 \hat{x}+\tau_2 \hat{y}),\\
	&G_n=(-1)^n4(g_n-\rho_i^2),\\
	&S_q^w(\mathbf{Q})=\frac{4}{N}\sum_{i,j}e^{i\mathbf{Q}(\mathbf{R}_i-\mathbf{R}_j)}w_iw_j,
	\end{aligned}
	\label{eq:parameter}
\end{equation}
where $\rho_i=\frac{N_f}{N_s}$ is the concentration of localized electrons.
$N_f$ and $N_s$ are the numbers of lattice sites and localized electrons, respectively.
Above-mentioned quantities reveal a charge density wave (CDW) state at low temperature around the half-filling situation. Away from half filling, inhomogeneous is indicated by the charge distribution in the real space, which has finite density of state around the Fermi energy.
As for high temperature, the phase is detected by the properties of transports, thermodynamics and the spectral function. Accordingly, we elaborate a phase diagram under strong coupling ($U=10$) on the $\mu-T$ plane with rich quantum states. The result is summarized in Fig.~\ref{fig:phase diagram}, where three different kinds of NFLs are uncovered with distinguishable thermodynamics and transports.
At high temperature, around the half filling correlation opens the gap and leads to the Mott insulator.
Doping holes into half-filled FK model drives the Mott insulator-metal transition accompanied with Mott quantum criticality.
With decreasing particle density, the Mott insulator crossovers to the first non-Fermi liquid (NFL-I) state, the strange metal state, the second non-Fermi liquid (NFL-II) state, subsequently.
Actually, the resistivity displays unambiguous bifurcating quantum scaling behavior
\begin{equation}
\rho(T,\delta \mu)=\rho^{\ast}(T)f(T/T_0(\delta \mu)),
 	\label{eq:quantum_scaling}
\end{equation}
where $T_0(\delta \mu)=c|\delta \mu|^{z\nu}$, $\delta \mu=\mu-\mu^{\ast}(T)$ and $\mu^{\ast}(T)$ is the critical 'zero field' trajectory corresponding to the 'separatrix' line.
$\rho^{\ast}(T)$ is calculated for $\mu=\mu^{\ast}(T)$ and $f(x)$ denotes the unknown scaling function.The resistivity on the separatrix is almost independent with temperature.
The NFL-I and the strange metal together compose the quantum critical region (see Fig.~\ref{fig:Cv}c).
The insulator-like branch and metallic branch satisfy different scaling functions $f$, but have the same critical exponent $z\nu=1.25$.
The NFL-I is near the Mott insulator and has insulator-like feature, where a upturn exists in the resistivity curves at low temperature.
The strange metal state with a lower electron density instead shows the metallic transport properties.
In such a quantum critical region, the FK model transits from insulator to metal.
As shown in Fig.~\ref{fig:Cv}(a,b), the strange metal state is confirmed with linear-$T$ resistivity and logarithm heat capacity coefficient dependence.
In metallic states, resistivity follows $\rho(T)=\rho_{0}+AT^{n}$.
Around the separatrix line of quantum critical region, resistivity shows a quadratic dependence of temperature ($n=2$), and turns to the linear dependence ($n=1$) when crossover to strange metal.
Further decreasing electron density, the scaling behavior is lost but linear-$T$ resistivity sustains, which is referred to NFL-II state.

\section{the violation of Luttinger theorem}\label{sec3}

\begin{figure}
	\centering
	\includegraphics[width=1\columnwidth]{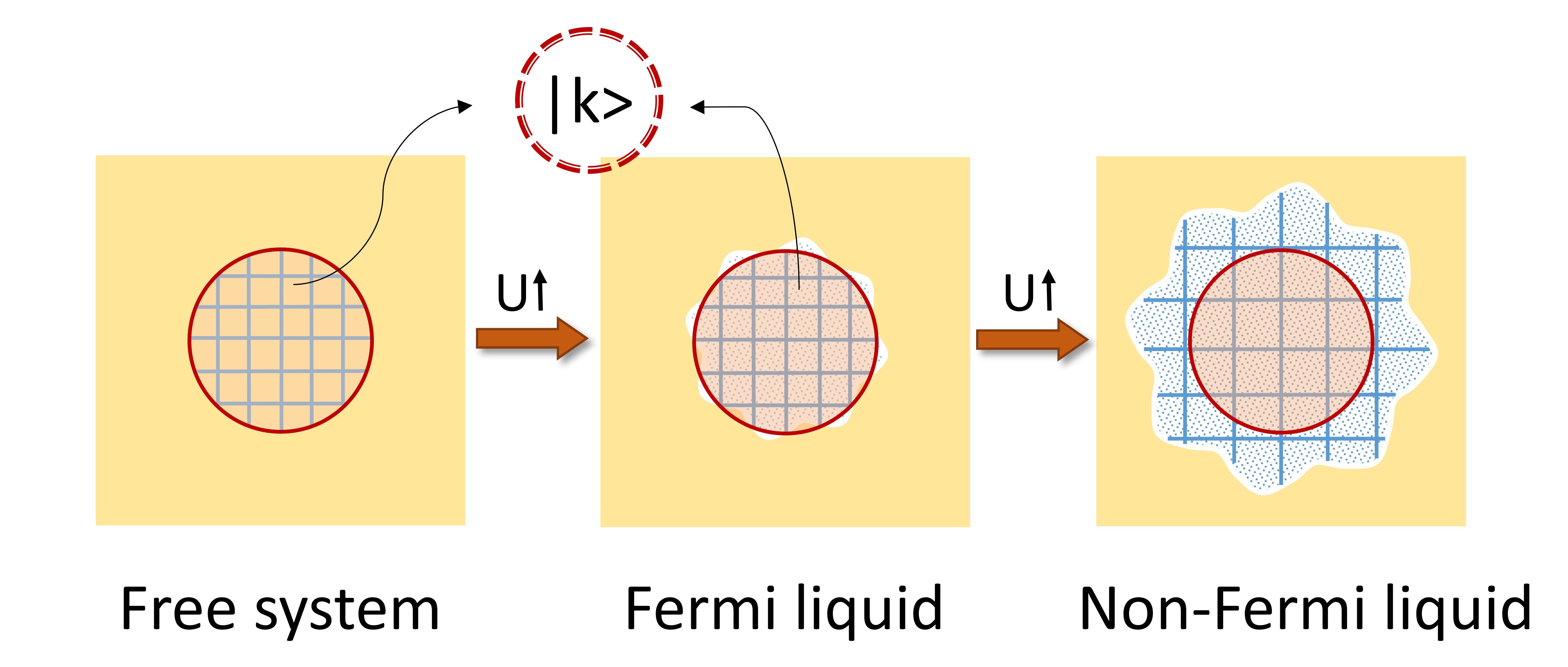}
	\caption{Schematic picture of the violation of Luttinger's theorem. For a free system with specific filling the volume of Fermi surface can be constructed by the Luttinger's theorem.		
		The red circle denotes the Fermi surface of a free system.
		The white region is enclosed by the Fermi surface of the interacting system.
		Adiabatically turning on the interaction in the free system (left) with fixed particle density, the volume enclosed by Fermi surface sustains as the free one until some critical coupling strength (middle).
		Under strong enough coupling (right), the interaction changes the volume enclosed by the Fermi surface from the free system.
    In fact, The violation of Luttinger's theorem suggests the NFL state has no quasiparticle which can be originated from the free system, we use the magnified grid in the Fermi surface to indicate changes of statistics. }
	\label{fig:LV3}
\end{figure}

Given the unconventional transports and thermodynamics, it is natural to explore the quasiparticle properties in the high temperature regime of the FK model.
 To achieve this aim, we examine the validity of the Luttinger's theorem\cite{PhysRevB.68.085113,PhysRev.118.1417,PhysRev.119.1153,PhysRevLett.84.3370},
which states that if Landau quasiparticle exists, the volume enclosed by Fermi surface is consistent with its density of particles.
Such theorem has been proved originally by Luttinger and Ward in terms of Luttinger-Ward functional in the framework of perturbation theory \cite{PhysRev.118.1417,PhysRev.119.1153}, and later by Oshikawa's non-perturbative topological argument \cite{PhysRevLett.84.3370}. It now has been accepted as a key feature of FL. Mathematically, the Luttinger's theorem defines the Luttinger integral ($\mathrm{LI}$) below
\begin{equation}
	\mathrm{LI}=\int_{\theta(\mathrm{Re}G(\textit{\textbf{k}},\omega=0))} \frac{d^dk}{(2\pi)^d},
	\label{eq:IL}
\end{equation}
and for general FL it must be equal to density of particles ($n_{c}$)
\begin{equation}
	\mathrm{LI}=n_{c}.
	\label{eq:LT}
\end{equation}
Here, $G(\textit{\textbf{k}},\omega)$ is the retarded single-electron Green function and $\theta(x)$ is the standard unit-step function with $\theta(x>0)=1$ and $\theta(x<0)=0$.

Intuitively, if we use the momentum space single-particle state $|\textit{\textbf{k}}\rangle$,
the particle density $n_c$ can be obtained by integrating the Fermi-Dirac distribution function with the weight $w(k)$ for each $|\textit{\textbf{k}}\rangle$
\begin{equation}
	\mathrm{n_c}=\int w(k) \times f(k)\frac{d^dk}{(2\pi)^d}.
	\label{eq:IL}
\end{equation}
At zero temperature in a free system, every $|\textit{\textbf{k}}\rangle$ state has the same spectral weight $w(k)=1$, and the Fermi distribution function $f(k)$ is the unit-step function of energy $f(k)=\theta(-\epsilon_\textit{\textbf{k}})$.
Thus, the particle density is equal to the proportion enclosed by the Fermi surface, which is the same as the Luttinger integral and naturally the Luttinger's theorem is satisfied.

Note that the Luttinger integral is the positive proportion of the real part of retard Green function in the first Brillouin zone, mathematically equal to the proportion enclosed by the Fermi surface.
Actually, for any system the Luttinger integral can be interpreted as the integral of the Fermi-Dirac distribution function with a constant weight $w(k)=1$
\begin{equation}
	\mathrm{LI}=\int 1 \times f(k)\frac{d^dk}{(2\pi)^d},
	\label{eq:IL2}
\end{equation}
which is the counting of occupied $|\textit{\textbf{k}}\rangle$ state.
According to Eq.~\ref{eq:IL} and Eq.~\ref{eq:IL2}, the working of the Luttinger's theorem suggests $w(k)=1$, i.e., in momentum space every volume element $\frac{d^dk}{(2\pi)^d}$ can contain 1 electron.
Thus the counting of electron number is equal to the counting of occupied $|\textit{\textbf{k}}\rangle$ state.
The many-body state and the excitation can be described clearly by the wave vector $\textit{\textbf{k}}$, which is the characteristic of the quasiparticle picture.

As shown in Fig.~\ref{fig:LV3}, by adiabatically turning on the correlation, the free fermion system evolves into FL, with $|\textit{\textbf{k}}\rangle$ state modified on the perturbative level.
In the FL state, the correspondence between the original state and the final state is valid, although the dispersion relation is modified by interaction.
The correspondence between free system and the FL suggests the presence of quasiparticle, where the weight of each $|\textit{\textbf{k}}\rangle$ state is still invariant as $w(k)=1$.
Actually, although the volume of new Fermi surface in FL is invariant, its shape should depend on the form of self-energy.
With an isotropic interaction, the real part of self-energy in momentum space is also isotropy, leading to an invariant Fermi surface.
Since the interaction in FK model is isotropy, in its FL regime we expect a Fermi surface being the same with free system.
However, with anisotropic interaction, the resultant self-energy is inhomogeneous in momentum space, and thus the Fermi surface can be deformed even within the FL(see the white region in the middle panel of Fig.~\ref{fig:LV3}).
Even though the interaction may change the occupied $|\textit{\textbf{k}}\rangle$ state, if only $w(k)=1$, the volume enclosed by Fermi surface is invariant with fixed particle density. It suggests the key feature of FL is the invariant spectral weight.

\begin{figure}
	\centering
	\includegraphics[width=1\columnwidth]{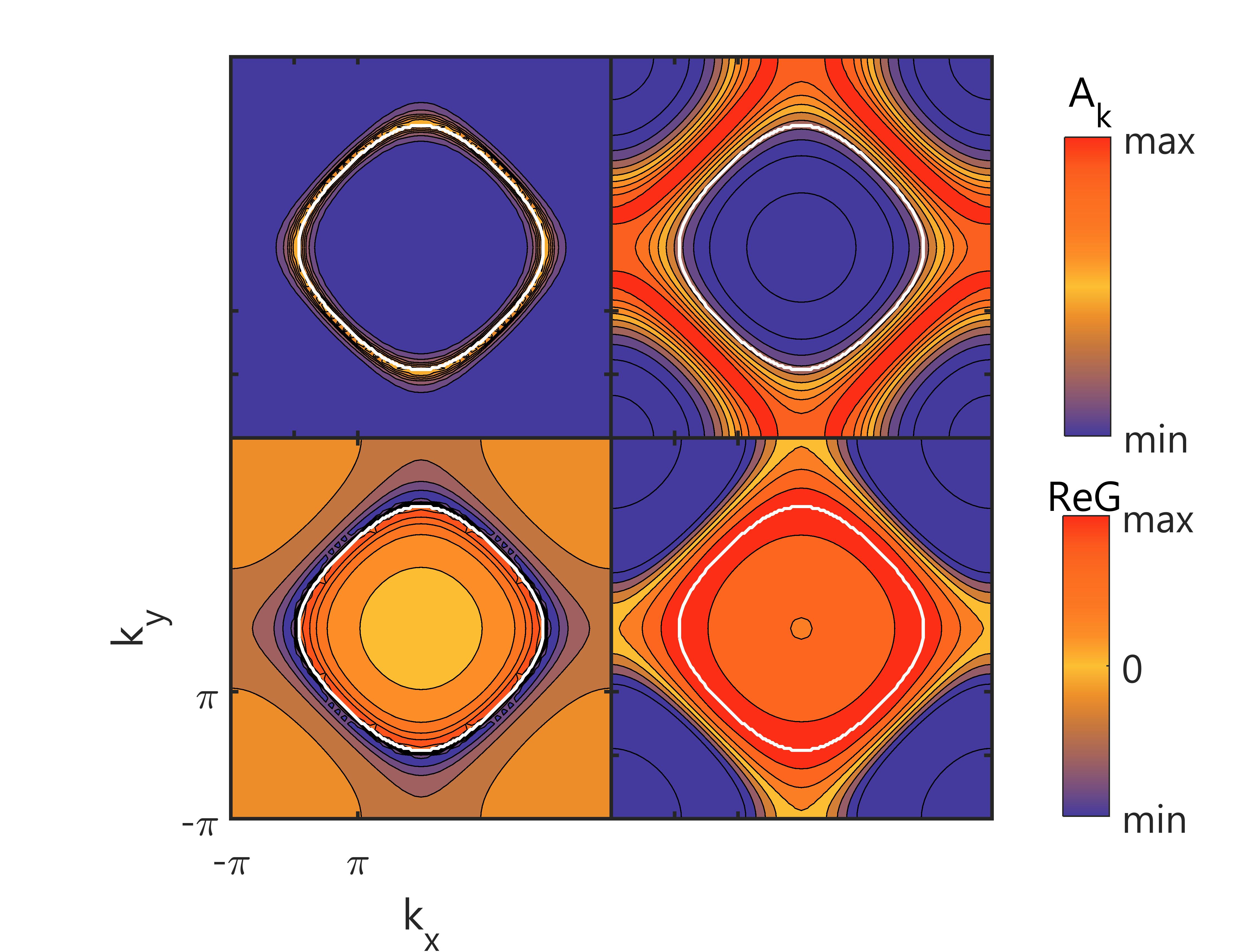}
	\caption{ Spectral function at Fermi energy $A_{\textit{\textbf{k}}}$$(\omega=0)$ (upper panel) and real part of Green function at Fermi energy $\mathrm{Re}G({\textit{\textbf{k}}},\omega=0)$ (lower panel) in the FL (left panel, $U=1$, $T=0.2$, $n_c=0.3$) and the SM state ($U=10$, $T=0.2$, $n_c=0.3$, left panel). According to the Luttinger's theorem, the Fermi surface of a free system with particle density $n_c=0.3$ is denoted by the white circle. The location of Fermi surface is in accordance with the zero values of the real part of Green function.}
	\label{fig:LV2}
\end{figure}

We emphasize that the Luttinger's theorem is valuable for study of exotic metallic states.
Since given a specific filling, the Luttinger's theorem provides a method to reconstruct the Fermi surface of a free system with the same particle density.
For a specific correlated system, we can confirm the presence of quasiparticle if its Fermi surface enclose the same volume with the free system.
Figure.~\ref{fig:LV2} displays the imagine and real part of the retard Green function of the FK model at $n_c=0.3$ under different coupling strength, respectively.
The upper panel plots $c$ electron's spectral function $A_{\textit{\textbf{k}}}(\omega=0)$ in the first Brillouin zone and the lower panel plots the real part of Green function $\mathrm{Re}G(\textit{\textbf{k}},\omega=0)$, both of which are plotted around the Fermi energy.
Based on the Luttinger's theorem, we utilize the particle density to plot the Fermi surface of a free system with the white line.
Note that no matter in the FL (left panel), or in the strange metal state (right panel), the location of Fermi surface is always coincident with the zero values of the real part of Green function.
It makes sure that in the strange metal state the Luttinger integral keeps as the proportion enclosed by the Fermi surface.

For a weak coupling situation (left panel, $U=1$), the interacting Fermi surface coincide with the free system, suggesting a FL nature.
As for the strong coupling situation (right panel, $U=10$), the interacting Fermi surface becomes heavily deformed.
The volume of Fermi surface is greater than the free one, which involves more possible occupied $|\textit{\textbf{k}}\rangle$ states.
As shown in the right panel of Fig.~\ref{fig:LV3}, when increasing the interaction, the correspondence between the original state in free system and the final many-body state in NFL disappears at some critical coupling strength.
This suggests that the average spectral weight decreases $\overline{w(k)}<1$
and averagely a $|\textit{\textbf{k}}\rangle$ state can contain electron less than 1.
For conventional fermions, the Pauli exclusion principle permits only one electron in a state,
whereas in the NFL the strong interaction effectively increases the exclusion in statistic.

\begin{figure}
	\centering
	\includegraphics[width=0.95\columnwidth]{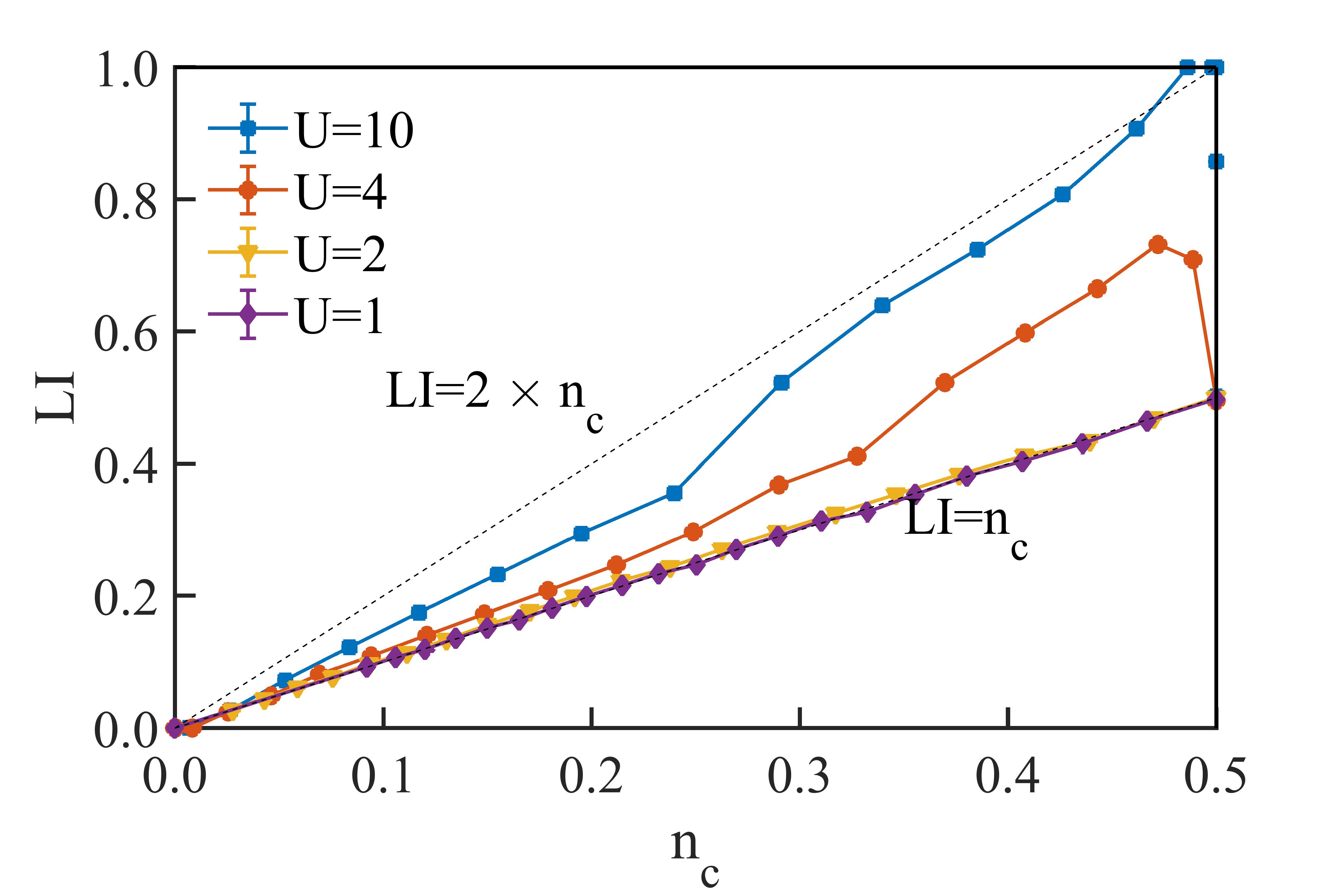}
	\caption{An overall spectacular of the violation of Luttinger's theorem. The Luttinger integral versus density of electron $n_c$ at $T=0.2$. In FL ($U=1$, $U=2$), the electron density and Fermi surface is in accordance with the Luttinger's theorem. At stronger coupling ($U=4$), the proportion of the first Brillouin zone enclosed by Fermi surface deviates from the density of electrons for most doping regime thus confirms their NFL nature. Further increasing coupling strength leads to a stronger deviation ($U=10$).}
	\label{fig:LV1}
\end{figure}

In Fig.~\ref{fig:LV1}, we demonstrate the overall Luttinger integral with varying coupling strength and varying filling.
As reference, two dotted black line is plotted as $n_c=\mathrm{IL}$ and $n_c=2 \times \mathrm{IL}$. Here, $n_c=\mathrm{IL}$ is the Luttinger integral of a free system, around which the Luttinger's theorem works well.
At exactly half filling, the Luttinger's theorem works well at any coupling strength, due to the existence of particle-hole symmetry.
With increasing coupling strength, the Luttinger's theorem sustains to $U=2$. When $U>2$ the violation emerges and the magnitude increases gradually with increasing interaction.
The degree of deviation corresponds to the degree of deformation of Fermi surface, compared with the free system.
Note that for a specific coupling strength $U$, the particle density changes the degree of deviation, but the deviation/identical of two Fermi surfaces is robust with doping. This robust violation corresponds to the rigid band structure of the FK model.
It suggests a robust NFL-like nature for all paramagnetic phases in the phase diagram (see Fig.~\ref{fig:phase diagram}).

Actually for the lattice model with finite sites, the above momentum integral should be replaced by discrete summation of the $k$-point grid. Frankly speaking, Eq.~\ref{eq:IL} is strictly valid only for zero temperature. Since interesting physics like NFLs and the strange metal behavior often exist at finite-$T$, here we follow the recent quantum determinant Monte Carlo study on the doped Hubbard model in Ref.~\onlinecite{PhysRevB.104.235122} and still use Eq.~\ref{eq:IL} to estimate the validity of the Luttinger's theorem at finite temperature. As indicated in our previous work, at high temperature regime the thermal effect on the Luttinger integral is small \cite{PhysRevB.104.165146}.
As shown in Fig.~\ref{fig:LV1}, at small filling Luttinger integral displays a linear dependence of the particle density $\text{IL}=a \times n_c$ whereas $a  \ne 1$ indicates the changes of statistics in NFLs.
Here, the spectral weight is $w(k)=\frac{1}{a}$.
The linear-$n_c$ behavior sustains to some critical particle density, which closely associated with the band structure.
The critical $n_c$ is correspondent to the coincidence of Fermi energy and the pseudo energy gap.
Not only the critical particle density ($n_c \sim 0.25$), but also the critical coupling strength ($U_c=2$) all turn out to be closely related to the band structure.
These properties will be further discussed in the binary disordered system in the final section.

\section{the Hubbard-I approximation and composite fermions analysis}\label{sec4}
\begin{figure}
	\centering
	\includegraphics[width=0.95\columnwidth]{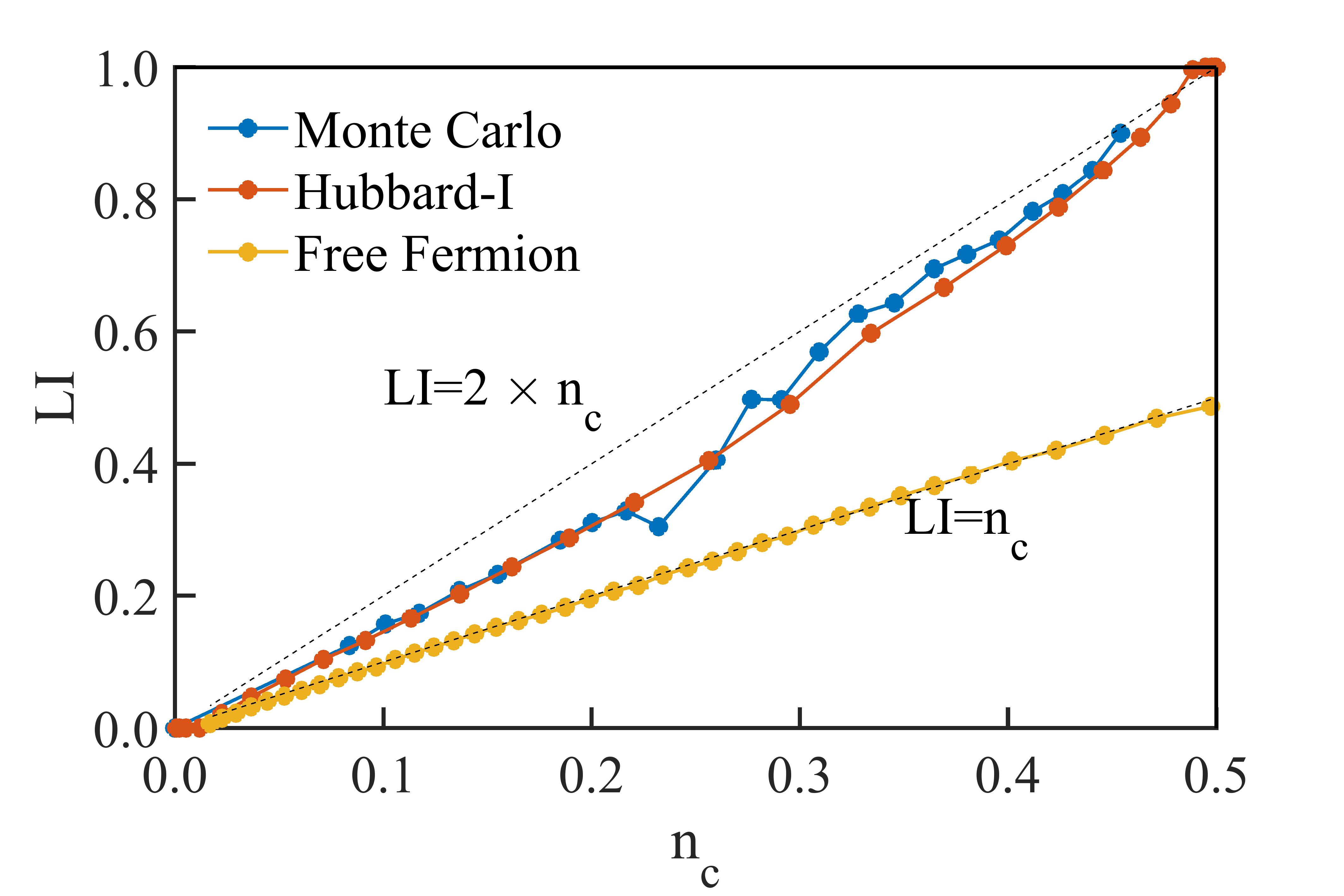}
	\caption{Luttinger integral (IL) versus density of electron $n_c$ at $T=0.2$. It displays the violation of the Luttinger's theorem by Monte Carlo simulation (blue, $U=10$) and the Hubbard-I approximation method (orange, $U=10$), respectively. The Luttinger integral of the free fermion system (yellow) is plotted as reference.}
	\label{fig:Hubbard-ReG}
\end{figure}

In the last section, with unbiased Monte Carlo simulation we study the violation of Luttinger's theorem in the FK lattice.
In this section, we want to further study the microscopic mechanism of the violation of the Luttinger's theorem analytically.
In the high temperature situation, one can follow the Hubbard-I approximation to give a rough solution for FK model.
At first, we use the equation of motion (EOM) formalism and define the retarded Green's function as \cite{doi:10.1098/rspa.1963.0204}
\begin{equation}
	G_{i,j,\sigma}(\omega)=\langle\langle \hat{c}_{i \sigma}|\hat{c}^\dagger_{j \sigma} \rangle\rangle.
	\label{eq:Hubbard-I1}
\end{equation}
By using the standard EOM relation,
\begin{equation}
	\omega \langle\langle \hat{A}|\hat{B}\rangle\rangle_{\omega}=\langle[\hat{A},\hat{B}]_+ \rangle+\langle \langle[\hat{A},\hat{H}]_-| \hat{B}\rangle\rangle_{\omega}.
	\label{eq:Hubbard-I2}
\end{equation}
For the FK model, the Hamiltonian can be rewritten as
\begin{equation}
	\hat{H}=-t\sum_{i,j}\hat{c}_{i}^{\dag}\hat{c}_{j}+
	\frac{U}{2}\sum_{i}(2\hat{w}_{i}-1)\hat{n}_{i}+(\frac{U}{2}-\mu)\sum_{i}\hat{n}_{i}.
	\label{eq:model2}
\end{equation}
At half-filled situation, it follows that
\begin{equation}
	\omega \langle\langle \hat{c}_{i}|\hat{c}^\dagger_{j }\rangle\rangle_{\omega}=\delta_{ij}-t \Delta_{im}\langle\langle \hat{c}_{m}|\hat{c}^\dagger_{j}\rangle\rangle_{\omega}+\frac{U}{2} \langle\langle (2\hat{w}_{i}-1)  \hat{c}_{i}|\hat{c}^\dagger_{j}\rangle\rangle_{\omega}.
	\label{eq:Hubbard-I3}
\end{equation}
Here, $\Delta_{im}$ denotes m is the nearest-neighbor site of i. For $\langle\langle (2\hat{w}_{i}-1)  \hat{c}_{i}|\hat{c}^\dagger_{j}\rangle\rangle_{\omega}$, we have
\begin{equation}
	\begin{aligned}
	\omega \langle\langle (2\hat{w}_{i}-1)  \hat{c}_{i }|\hat{c}^\dagger_{j}\rangle\rangle_{\omega}= & \langle (2\hat{w}_{i}-1) \rangle \delta_{ij} 
	-t \Delta_{il} \langle\langle (2\hat{w}_{i}-1) \hat{c}_{l}|\hat{c}^\dagger_{j}\rangle\rangle_{\omega} \\
	&+\frac{U}{2}   \langle\langle \hat{c}_{i }|\hat{c}^\dagger_{j }\rangle\rangle_{\omega}.
	\end{aligned}
	\label{eq:Hubbard-I4}
\end{equation}
If no further EOM is involved, to close EOM we have to decouple $\langle\langle (2\hat{w}_{i}-1)  \hat{c}_{l}|\hat{c}^\dagger_{j }\rangle\rangle_{\omega}$ as
\begin{equation}
	\langle\langle (2\hat{w}_{i}-1)  \hat{c}_{l}|\hat{c}^\dagger_{j}\rangle\rangle_{\omega} \simeq \langle 2\hat{w}_{i}-1\rangle \langle\langle   \hat{c}_{l}|\hat{c}^\dagger_{j}\rangle\rangle_{\omega}.
	\label{eq:Hubbard-I5}
\end{equation}
In paramagnetic strong coupling regime, there is no finite CDW order, thus $\langle(2\hat{w}_{i}-1)\rangle=0$. Meanwhile, the contribution from $\langle\langle (2\hat{w}_{i}-1)  \hat{c}_{i}|\hat{c}^\dagger_{j}\rangle\rangle_{\omega}$ vanishes due to the decoupling and the above equation has a complete solution:
\begin{equation}
	(\omega-\frac{U^2}{4\omega}) \langle\langle \hat{c}_{i}|\hat{c}^\dagger_{j}\rangle\rangle_{\omega}=\delta_{ij}-t \Delta_{im}\langle\langle \hat{c}_{m}|\hat{c}^\dagger_{j}\rangle\rangle_{\omega},
	\label{eq:Hubbard-I6}
\end{equation}
which can be written as
\begin{equation}
	(\omega-\frac{U^2}{4\omega}) G_{i,j}(\omega)=\delta_{ij}-t \Delta_{im}G_{m,j}(\omega).
	\label{eq:Hubbard-I7}
\end{equation}
Now, performing the Fourier transformation
\begin{equation}
	G_{i,j}(\omega)=\frac{1}{N_s} \sum_k e^{ik(R_i-R_j)}G(k,\omega),
	\label{eq:Hubbard-I8}
\end{equation}
we have
\begin{equation}
	\begin{aligned}
		&\sum_k (\omega-\frac{J^2}{16\omega}) G(k,\omega) e^{ik(R_i-R_j)}=\sum_k e^{ik(R_i-R_j)} \\
		&-t \Delta_{im}G(k,\omega) e^{ik(R_m-R_j+R_i-R_i)}.
	\end{aligned}
	\label{eq:Hubbard-I9}
\end{equation}
Here, $-t \Delta_{im}e^{ik(R_m-R_i)}=-t\sum_{\delta} e^{-ik\delta}=\varepsilon_k$ and we have $\sum_k[(\omega-\frac{U^2}{4\omega})G(k,\omega)-1]e^{ik(R_i-R_j)}=0 $, which gives the single-particle Green's function as
\begin{equation}
	\begin{aligned}
		G(k,\omega)=\frac{1}{\omega-\frac{U^2}{4\omega}-\varepsilon_k}=\frac{\alpha_k^2}{\omega-\hat{E}_k^+}+\frac{1-\alpha_k^2}{\omega-\hat{E}_k^-}.
	\end{aligned}
	\label{eq:Hubbard-I10}
\end{equation}
Here, the coherent factor $\alpha_k^2=\frac{1}{2} \left( 1+\frac{\varepsilon_k}{\sqrt{\varepsilon_k^2+U^2}} \right)$ and
\begin{equation}
	\begin{aligned}
		\hat{E}_k^{\pm}=\frac{1}{2} \left[ \varepsilon_k\pm \sqrt{\varepsilon_k^2+U^2} \right].
	\end{aligned}
	\label{eq:Hubbard-I11}
\end{equation}
With above single-particle Green's function, we plot the Luttinger integral at $U=10$ by the Hubbard-I approximation (orange line with circle marker) as a function of particle density and compare it with the Monte Carlo result (blue line with circle marker) in Fig.~\ref{fig:Hubbard-ReG}.
As reference, the Luttinger integral of the free fermion system is plotted in yellow line with circle marker.

It turns out that the Hubbard-I approximation provides an accurate evaluation for the Fermi surface at strong coupling regime. Compared with the Hubbard model, in the FK model Hubbard-I approximation works better, which can almost recover the violation of Luttinger's theorem.
This advantage in the FK model is attributed to the commutation between the interaction and hopping terms
\begin{equation}
 [2\hat{w}_{i}-1,\hat{c}_{i}\hat{c}^\dagger_{j }]=0,
	\label{eq:commuation}
\end{equation}
where the EOM truncates naturally at the second order only if we assume Eq.~\ref{eq:Hubbard-I5} works.

As mentioned above, the violation of Luttinger's theorem suggests the correspondence between FL and NFL is lost.
We cannot access to an approximated formalism of the quasiparticle by renormalizing parameters, such as the effective mass.
This remind us that an unconventional type of quasiparticle (not Landau quasiparticle) excitation should be expected, if exists.
To access an intuitive understanding of the violation of the Luttinger's theorem in terms of approximated quasiparticle, we provide a composite fermion approach with similar result compared with Hubbard-I approximation.
We note that the interaction term in Eq.~\ref{eq:model1} can be interpreted as the hybridization between the itinerant electrons and some composite fermions $f^{\dagger}_j=(2\hat{w}_j-1)c^{\dagger}_j$, which satisfies the commutation relation of fermions
\begin{equation}
	\{f^{\dagger}_i,f_j\}=(2\hat{w}_i-1)(2\hat{w}_j-1)\{c^{\dagger}_i,c_j\}=\delta_{ij}.
	\label{eq:commutation}
\end{equation}
However, the commutation relation between the composite fermion and the itinerant electron is nontrivial
 \begin{equation}
 	\{c^{\dagger}_i,f_j\}=(2\hat{w}_j-1)\{c^{\dagger}_i,c_j\}=(2\hat{w}_j-1)\delta_{ij}.
 	\label{eq:commutation2}
 \end{equation}
Now the FK Hamiltonian can be written as
\begin{equation}
	\hat{H}=-t\sum_{i,j}\hat{c}_{i}^{\dag}\hat{c}_{j}+
	\frac{U}{4}\sum_{i}(\hat{f}^{\dagger}_{i}\hat{c}_{i}+\hat{c}^{\dagger}_{i}\hat{f}_{i})+(\frac{U}{2}-\mu)\sum_{i}\hat{n}_{i}.
	\label{eq:model3}
\end{equation}
We assume $\{c^{\dagger}_i,f_j\}$ can be replaced by its average value. At the high temperature paramagnetic regime $<(2\hat{n}_j-1)>=0$ and thus the composite fermion and itinerant electron restore a standard commutation relation $\{c^{\dagger}_i,f_j\}=0$.
It suggests at high temperature we can treat the composite fermion as conventional fermion.
By Fourier transformation, the Hamiltonian can be written as
\begin{equation}
	\hat{H}=\sum_{k}
	 \left(                 
	\begin{array}{cc}   
		\hat{c}_{k}^{\dag} & \hat{f}_{k}^{\dag}\\  
	\end{array}
	\right)
	 \left(                 
	\begin{array}{cc}   
		\epsilon_k+\frac{U}{2}-\mu & \frac{U}{4} \\  
		\frac{U}{4} & 0 \\  
	\end{array}
	\right)
	 \left(                 
	\begin{array}{c}   
		\hat{c}_{k} \\  
		\hat{f}_{k} \\  
	\end{array}
	\right).
	\label{eq:model3_1}
\end{equation}
The dispersion is given by diagonalizing the Hamiltonian
\begin{equation}
	\begin{aligned}
		\hat{E}_k^{\pm}=\frac{1}{2} \left[ \varepsilon_k\pm \sqrt{\varepsilon_k^2+\frac{U^2}{4}} \right].
	\end{aligned}
	\label{eq:Hubbard-I11}
\end{equation}
 This dispersion is similar with the Hubbard-I results, with a gap $\triangle E=\frac{U}{2}$ smaller than $\triangle E_{Hubbard-I}=U$.
This composite fermion phenomenologically suggests that the elementary excitation of the FK model consists of the contribution of both itinerant electron and also the composite fermion.
Therefore, it is hard to describe the many-body state in terms of a single (Landau) quasiparticle picture.
As particle number increasing by 1, in the momentum space the correspondent many-body state occupied a volume larger than $\frac{d^dk}{(2\pi)^d}$.


\section{Discussion and Conclusion}\label{sec5}
\subsection{Phase diagram at half filling}\label{sec5_1}

\begin{figure}
	\centering
	\includegraphics[width=0.95\columnwidth]{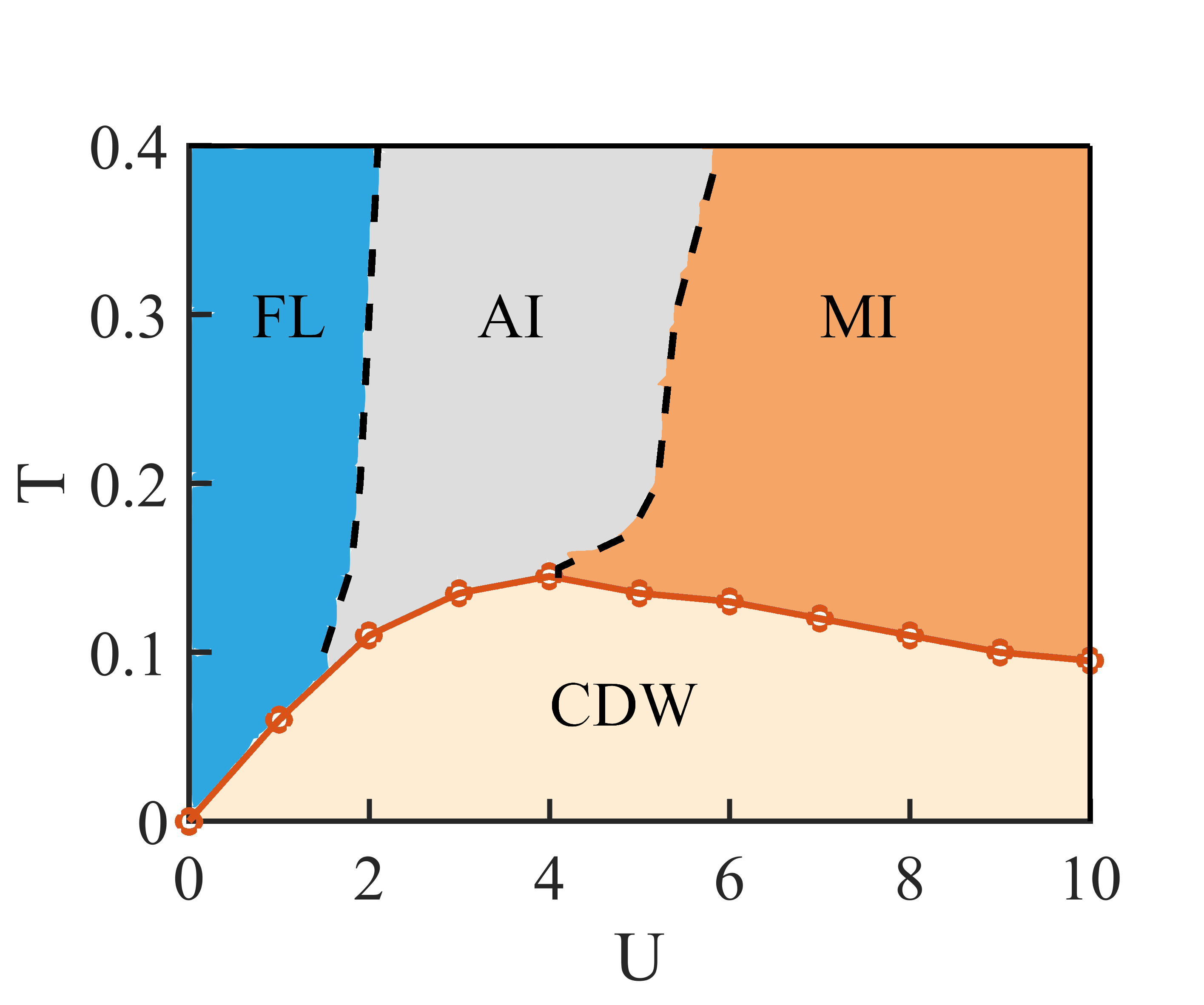}
	\caption{The phase diagram of the FK model in the $U-T$ plane at half-filling situation. The low temperature regime is a CDW state under any finite coupling. At high temperature three different states are uncovered. Under weak coupling system sustains as a FL until $U=2$. At moderate coupling, the Anderson insulator (AI) emerges with a crossover. Further increasing correlation leads to the MI with full opened energy gap.}
	\label{fig:phase diagram2}
\end{figure}

We notice the robust violation/working of the Luttinger's theorem for a specific coupling strength in the doped FK model, which is invariant under different doping.
Therefore, the Luttinger's theorem can not only help tell NFLs from FLs in the doped system,
but also help distinguish different state under different coupling strength at half filling,
although the working of Luttinger's theorem is protected there by the particle-hole symmetry.
According to the order parameter in Eq.~\ref{eq:parameter}, the density of state, the inverse participation ratio,
and the critical coupling strength where the Luttinger's theorem fails with doping, the half-filling phase diagram of the FK model in the $U-T$ plane can be elaborated as Fig.~\ref{fig:phase diagram2}.
At low temperature the CDW state exists under any finite interaction $U$.
At high temperature, under strong coupling ($U>6$), the FK model is a Mott insulator with fully opened energy gap.
Under weak coupling ($U<2$), the FL nature sustains.
Under a moderate coupling ($2<U<6$), there exists finite density of state around Fermi energy while the two-band structure emerges.
The localization revealed by inverse participation ratio suggests an Anderson insulator.

Note that the phase diagram is different from the previous one in Ref.~\onlinecite{PhysRevLett.117.146601}.
In Ref.~\onlinecite{PhysRevLett.117.146601}, the whole metallic regime is indicated as the localization state, where the Anderson insulator state can be extrapolated to any finite small coupling at thermodynamic limit.
In this work, according to the critical coupling strength where the Luttinger's theorem is violated in the doped FK lattice, we find that the Anderson insulator state can be further divided into a FL state at weak coupling and a NFL state at strong coupling.
What's more, the FL state is not simply the effect of finite size, which can sustain to $U \sim 2$ at thermodynamic limit.

\subsection{A binary disordered system}\label{sec5_2}
Since the Anderson localization in FK model has long been attributed to the disorder effect \cite{PhysRevB.104.045116}, it is valuable to check the pure effect of disorder on the working of Luttinger's theorem.
We wonder whether increasing the strength of disorder will leads to the disappearance of quasiparticle.
Since the doped FK model possesses robust rigid two-band structure at strong coupling, some connection may exist between the robust violation of the Luttinger's theorem and the robust two-band structure, which also corresponds to the singularity of Green's function at high energy regime \cite{Heath_2020}.
However, we wonder whether the origin of band splitting is also important.
To this end, it is necessary to study a disordered system, which could induce a two-band structure at the absence of interaction.
Thus, in this section, we construct a Hamiltonian with interaction replaced by discrete binary disorder to study the Luttinger's theorem in a two-band system, which is written as:
\begin{equation}
	\hat{H}=-t\sum_{i,j}\hat{c}_{i}^{\dag}\hat{c}_{j}-(\mu_i-\mu_0)\sum_{i}\hat{n}_{i}.
	\label{eq:model4}
\end{equation}
Different from the FK model, where $\mu$ is fixed during a Monte Carlo simulation, in this disordered system chemical potential is composed by two part and the binary disorder is introduced by $\mu_i$.
For each site $i$, $\mu_i$ is randomly taken as $\mu_i=\pm {U}_\text{disorder}$. $\mu_0$ instead is a fixed number in Monte Carlo simulation and it is used to tuning the particle density.

\begin{figure}
	\centering
	\includegraphics[width=1.\columnwidth]{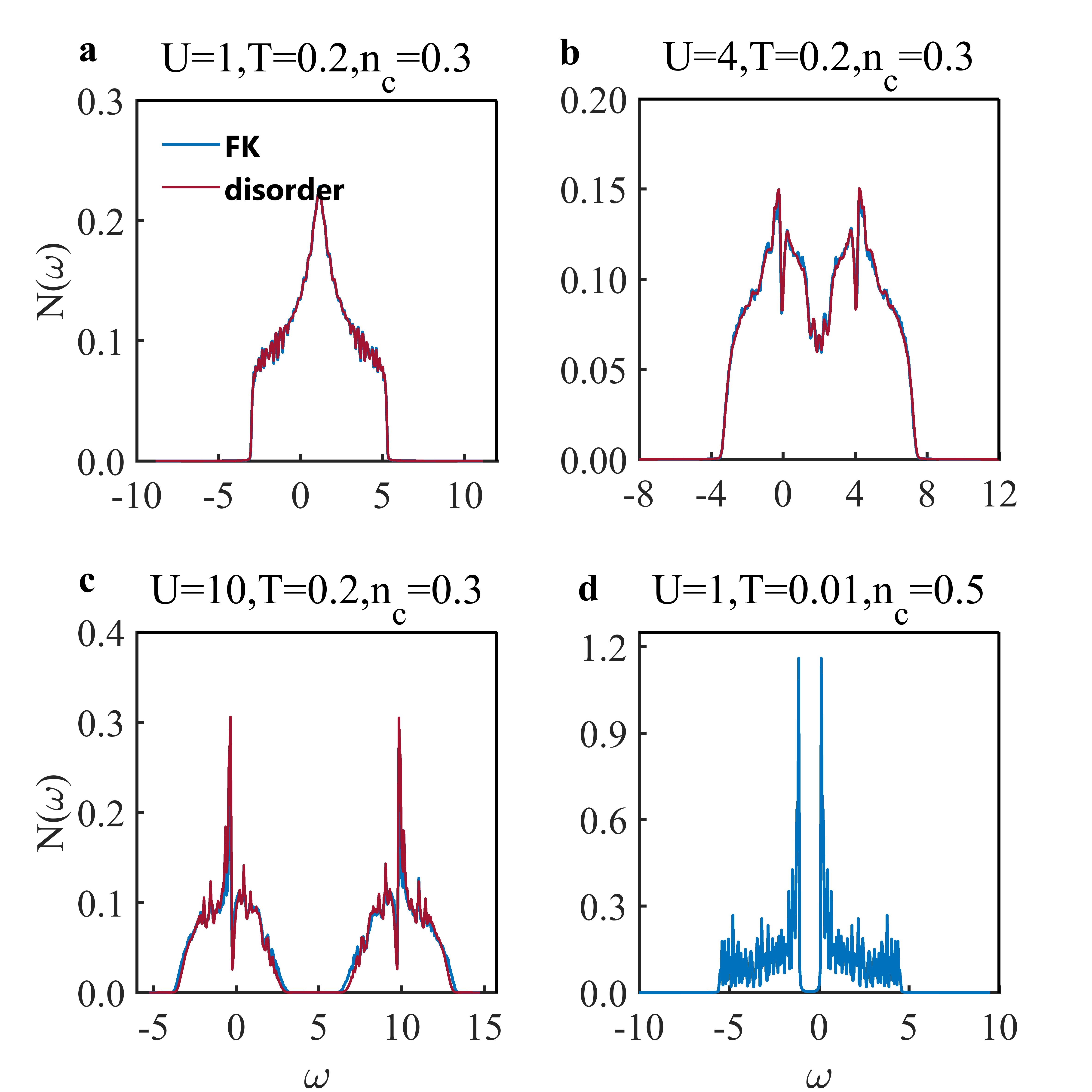}
	\caption{(a-c) Density of state under different coupling/disorder strength at $n_c=0.3$ at high temperature $T=0.2$. The density of state of the FK model is denoted by the blue line, and the binary disordered system is denoted by the red line. (d) In FK model the energy gap is opened under small interaction $U=1$ at low temperature $T=0.01$. This opened gap can easily be erased by either thermal fluctuation or doping holes.}
	\label{fig:dis_dos}
\end{figure}

We show the density of state at $n_c=0.3$ for this disordered system (red line) with different disorder strength in Fig.~\ref{fig:dis_dos}(a-c) and compare it with the FK model (blue line).
It turns out that when $|\mu_i|=U_{FK}$, $\mu_0=\mu_{FK}-\frac{U_{FK}}{2}$, the spectral function of the FK model agrees with the disordered system perfectly with varying coupling/disorder strength $U$. Even the location of pseudo gap is in accordance with each other.
At small coupling/disorder strength ($U=1$), only one quasiparticle peak exists and it is the same with the free electron situation.
Under a moderate coupling strength ($2<U<6$), the two-peak structure emerges while the gap is still not fully opened. Finite density of state around the Fermi energy and the two-peak structure leads to the localization of the FK/disordered system. This region is referred to as the Anderson insulator. Increasing coupling/disorder strength makes the gap fully opened, and the scale can be tuned by the magnitude of $U$.

\begin{figure}
	\centering
	\includegraphics[width=0.95\columnwidth]{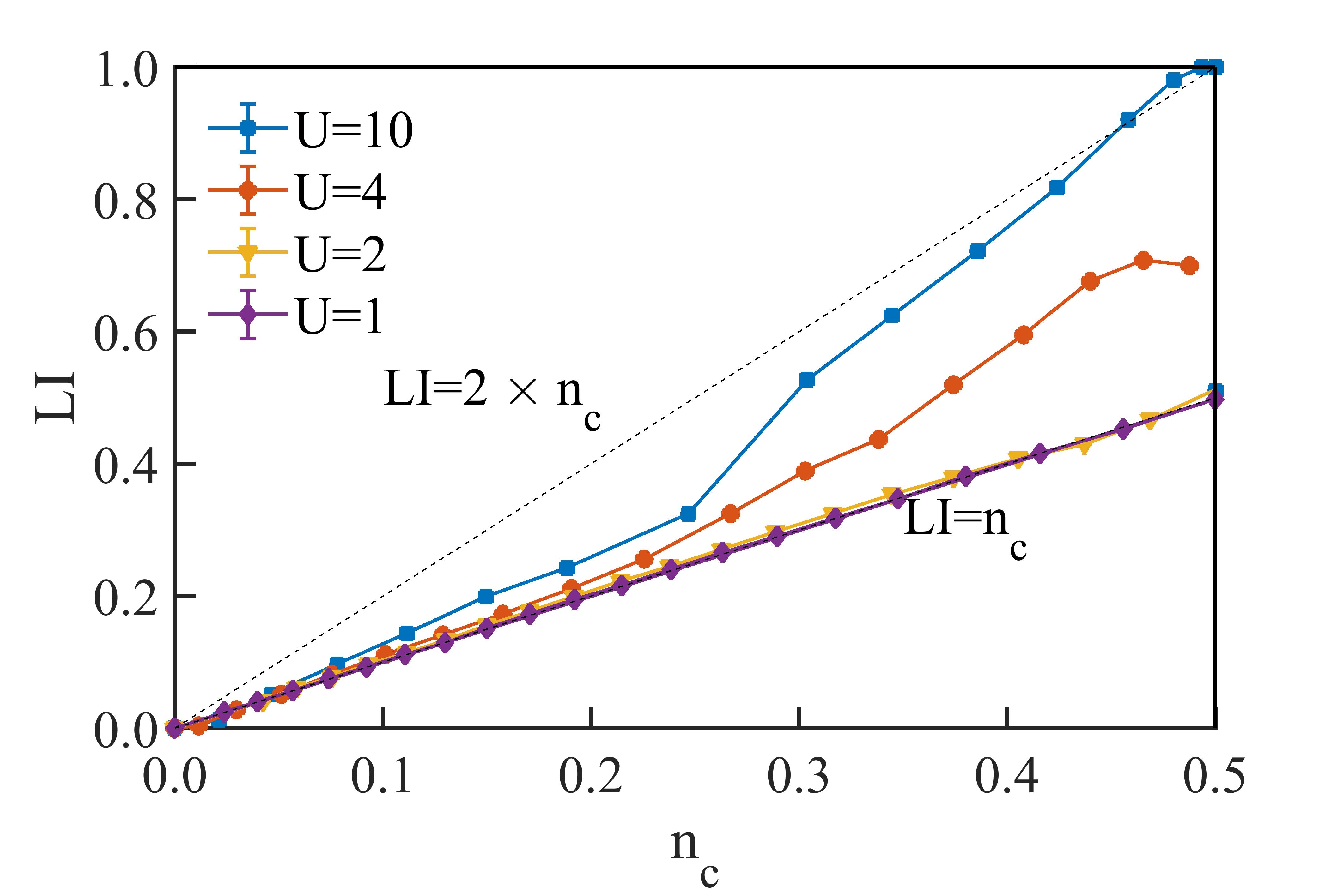}
	\caption{Luttinger integral (IL) versus density of electron $n_c$ at $T=0.2$. It displays the violation of the Luttinger's theorem in the binary-disordered system. Within a weak disorder (${U}_\text{disorder}=1$, ${U}_\text{disorder}=2$), system behaves as a general FL, where the electron density is in accordance with the Luttinger's theorem. Within stronger disorder (${U}_\text{disorder}=4$), the doped Anderson insulator show the NFL nature, where the proportion enclosed by Fermi surface deviates from the density of electrons for most doping regime. Further increasing disorder strength leads to a stronger deviation (${U}_\text{disorder}=10$). }
	\label{fig:4}
\end{figure}

As shown in Fig~.\ref{fig:4}, we further demonstrate the Luttinger integral in the disordered system, as a function of disorder strength and the particle density.
We study the working/violation of the Luttinger's theorem and the evolution of spectral function at different parameter, to confirm the role that band structure plays in the violation of the Luttinger's theorem.
As shown in Fig~.\ref{fig:dis_dos}a, when the one-band structure sustains ($U=1$), the Luttinger's theorem works well where the Fermi surface enclosing a volume consistent with the particle density $n_c$ (see Fig.~\ref{fig:4}).
Increasing the disorder ($U=4$) opens an energy gap $\triangle E \sim U$ in the middle of the spectrum and introduces the deviation of Luttinger integral, suggesting a deformed Fermi surface with larger enclosed volume.
Note that at $U=4$ the FK model is in the Anderson insulator state, where the gap is not fully opened.
It indicates that the violation of the Luttinger's theorem is directly caused by the two-peak structure, i.e, by the singularities of the self-energy at both high and low frequencies, rather than by the
fully opened Mott gap.
Further increasing the disorder ($U=10$) leads to greater deviation and finally accesses to the Mott insulator at half filling.

The Luttinger integral in Fig.~\ref{fig:4} is similar with the FK one shown in Fig.~\ref{fig:LV1}.
In both FK model and the disordered system, the working of Luttinger's theorem depends on the band structure.
We want to emphasize that if the Anderson localization is introduced by the Gaussian-type disorder instead of this discrete disorder,
the Luttinger's theorem will not be violated due to the absence of two-band structure.
Therefore, we conclude the NFL nature is not caused by localization.

As a function of particle density,
the behavior of Luttinger integral is associated closely with the characters of the spectrum function, e.g., another singularity around the pseudo gap.
In both $U=10$ and $U=4$ situations, the Luttinger integral displays a linear dependence of particle density when $n_c < 0.25$, where $n_c = 0.25$ is correspondent to the filling around pseudo gap.

\subsection{Some discussions}\label{sec5_3}

We note that previous analytical studies have demonstrated that two-band structure of the Hubbard and related models guarantees a violation of the Luttinger's theorem, 
where the violation stems from the zeros of the single-particle Green function which arises from the separation of scales of the upper and lower Hubbard bands \cite{PhysRevB.75.104503,PhysRevLett.110.090403,Phillips_2020}.
Actually, the proof is also available for the FK model.
In the Mott insulator state, at Fermi energy the self-energy diverges at $\epsilon_k=0$ in the momentum space, leading to the zeros of the single-particle Green function and a fully opened gap.
The divergence of self-energy makes Luttinger-Ward functional ill-defined. Slightly doping away from the half-filling situation, the Luttinger integral counts the zeros instead of poles as the particle density.
As we know, the zeros in momentum space corresponds to the Luttinger surface.
At half filling, the Luttinger surface just locates at the Fermi surface of a half-filling free system.
In fact, with any finite doping, in the FK model the Luttinger surface always exists at a specific energy ($\omega=\mu$).
However, when doping driven the FK model crossover from the Mott insulator to the NFL-I, the energy scale $\omega=\mu$ is too high to dominate the physics.
In the NFL-I, strange metal, and NFL-II, the Fermi surface instead emerges around Fermi energy.

The previous study based on cluster dynamical mean field theory suggests that the FK model is never a strict Landau FL even at weak coupling situation ($U/t \sim 0.5$) \cite{PhysRevB.94.081115}.
However, our work indicates that the Luttinger's theorem is working perfectly at $U/t=1$ as shown in the Fig.~\ref{fig:LV2}, which instead is a clear signal for the presence of quasiparticle.
The disagreement originates from the fact that the NFLs we discussed here are embedded in high temperature paramagnetic regime, where the effect of thermal fluctuation plays an important role.
Actually, as shown in Fig~.\ref{fig:dis_dos}d, CDW ground state of the FK model is robust for any finite $U$, in which the interaction splits the energy band and leads to fully opened gap.
However, different from the real NFLs, this two-band structure is not robust, which can be simply destroyed by either thermal fluctuation or doping holes.
Therefore, only conventional FL is revealed under weak coupling.
Only if the coupling strength is larger than the critical value $U_c$ ($U_c=2$) can the energy gap sustain under thermal fluctuation and doping.

\subsection{Conclusion}\label{sec5_4}

In this paper, with unbiased Monte Carlo simulation, we reveal an unambiguously violation of the Luttinger's theorem in the doped FK model.
Under strong coupling, doping holes in the half-filling FK lattice leads to a Mott insulator-metal transition.
Intriguingly, the NFL nature is robust under strong coupling at any finite doping.
 As shown in the Fig.~\ref{fig:phase diagram},
at high-temperature regime three different NFLs are revealed.
Empirically, we used to distinguish the NFL features by unconventional transport and thermodynamic properties,
such as the linear-$T$ resistivity and logarithm dependence of heat capacity coefficient in the strange metal state.
Although demonstrating various thermodynamical and transport behaviors, their NFL-like nature do originate from a common ground in spectrum.
Here, with the Luttinger's theorem, we demonstrate the deformation of Fermi surface in the NFLs is associated with the transfer of spectral weight.
After reliable numerical simulation, we try to analytically access to the violation of the Luttinger's theorem.
With the Hubbard-I approximation approach where we have solved the EOM to the second order,
the violation of the Luttinger's theorem is reproduced quantitatively.
To provide a more intuitive picture, we construct the composite fermion. It turns out the excitation in the FK model is
mixed of the itinerant electron and the composite fermion, which cannot simply described by elementary excitation formalism of a specific particle.
Considering the lack of quasiparticle, the breakdown of FL paradigm is predictable.
Finally, a binary disordered model is constructed to discuss the connection between the violation of Luttinger's theorem and the feature of spectrum.
This disordered system can cover the density of states in FK model perfectly.
In such a different model which emphasizes the disorder effect rather than interaction,
the two-band structure can also deform the Fermi surface directly.
As a conclusion, the violation of Luttinger's theorem in the doped Mott insulator is directly connected to the two-band structure, where the change of statistics can be taken as the key feature of NFLs underlying unconventional phenomena.

	\begin{acknowledgements}
This research was supported in part by Supercomputing Center of
Lanzhou University and NSFC under Grant No.~$11834005$, No.~$11874188$.	
We thank the Supercomputing Center of Lanzhou University for allocation of CPU time.		
	\end{acknowledgements}

\bibliography{ref}



\end{document}